\documentclass[pre,onecolumn,12pt]{revtex4}
\usepackage{epsfig}
\usepackage{graphicx}
\usepackage{epsfig}
\usepackage{amssymb}
\usepackage{amsmath}
\usepackage{graphicx}
\usepackage{hyperref}
\usepackage{color}

\normalsize

\definecolor{darkred}{rgb}{0,0,0.5}
\definecolor{darkgreen}{rgb}{0,0,0.5}
\definecolor{darkblue}{rgb}{0,0,0.5}

\begin{document}

\title{Mean-field 'Temperature' in Far From Equilibrium Systems}
\author{I. Santamar\'{\i}a-Holek\dag,A. P\'{e}rez-Madrid\ddag}

\affiliation{\dag UMJ-Facultad de Ciencias, Universidad Nacional Aut\'onoma de M\'exico,
Campus Juriquilla, Boulevard Juriquilla, No. 3001, C.P. 76230, Quer\'etaro,
M\'exico}
\affiliation{\ddag Departament de F\'{\i}sica Fonamental, Facultat de F\'{\i}sica,
Universitat de Barcelona, Av. Diagonal 647, 08028 Barcelona, Spain}


\vspace{2.5cm}
\begin{abstract}
We calculate the nonequilibrium mean-field 'temperature' of a Brownian system
in contact with a heat bath. We consider two different cases: an equilibrium
bath in the presence of strong external forces and a nonequilibrium bath. By
proving the existence of a generalized fluctuation-dissipation relation this
mean-field 'temperature' can be used to describe a nonequilibrium system as is
if it were in thermal equilibrium with a thermal bath at the 
 mean-field 'temperature' mentioned above. We apply our results to chemical reactions
in the presence of external forces showing how chemical
equilibrium and 
Kramers  rate constants are modified by the presence of these forces. \newline
\end{abstract}
\maketitle

\section{Introduction}

In the thermodynamic analysis of mesoscopic and macroscopic systems 
out of equilibrium an interesting question arises: Can we define a
nonequilibrium 'temperature'? Although this is possible, this 'temperature' cannot
be a thermodynamic temperature since there is not a thermodynamic zero principle behind
it. The consequences of this fact have been studied in detail in Ref.\cite{hernandez}.
Nevertheless, the concept of a nonequilibrium 'temperature' can be used to
parametrize the  quasi-equilibrium states of a nonequilibrium system 
\cite{hernandez},\cite{chetrite},\cite{berthier},\cite{granular},\cite{berthier2}.

The notion of a nonequilibrium 'temperature' can be made more precise by
noting that this is a statistical concept related to the energy of
a system involved in the erratic motion of its Brownian degrees of
freedom, that is, related to the thermal energy of the system. Since the
thermal energy of equilibrium and nonequilibrium systems will in general differ, one expects that the nonequilibrium
'temperature' of an out of equilibrium system will be different from that of the heat bath. Introducing the
nonequilibrium 'temperature' has the advantage that one may describe the
system as is if it were at thermal equilibrium with a hypothetical bath
with a temperature corresponding to this nonequilibrium 'temperature'. This
possibility has been studied for example for quantum corrections to the low
temperature of thermodynamic systems or thermal radiation not in equilibrium
in Ref. \cite{landau} (page 104, section 34 and page 189, section 63, respectively)
and in the case of granular matter in Ref. \cite{granular}. Here, by using
the nonequilibrium 'temperature', we will show that in
the case of chemical reactions an external force shifts  chemical
equilibrium and increases the reaction rates. This
effect may be particularly important in the case of photochemistry
where the incident light increases the velocity of the reaction.

In nonequilibrium systems such as small, confined or glass-like systems \cite%
{haxton}, introducing a nonequilibrium 'temperature' is appropriate 
 since this concept  is a consequence of the existence of internal constraints or
long-range forces and correlations which maintain the system out of
thermodynamic equilibrium. Some of these constraints may appear at low
temperatures and for small masses as in the case of quantum effects \cite%
{landau} mentioned above. In other cases, it is a purely classical effect when the range of
the interactions is similar to the size of the system.

In this article, we propose a definition of the nonequilibrium
'temperature' $T(a,t)$ in \textit{analogy} with the equipartition theorem 
\begin{equation}
k_{B}T(a,t)\rho (a,t)\equiv \int b^{2}f(b,a,t)db,  \label{zero}
\end{equation}%
where $k_{B}$ is Boltzmann's constant, $f(b,a,t)$ is a probability
distribution describing the state of the system which, for
convenience's sake, we assume to depend on a pair of conjugated slow $a$ and
fast $b$ variables (a velocity). In addition, $\rho (a,t)=\int f(b,a,t)db$ is a
reduced probability distribution. One of the goals of this article consists
of showing how this nonequilibrium 'temperature', which we will 
call effective 'temperature', is connected to the bath temperature 
and the forces which maintain the system out of equilibrium \cite{berthier}. We also analyze how the fluctuation-dissipation theorem (FDT)
is modified in systems far from equilibrium and how the effective
'temperature' comes into play in order to extend the  validity of this FDT to 
quasistationary nonequilibrium systems.

The paper is organized as follows: In section 2 we analyze Brownian motion
in a field of force. We obtain the Fokker-Planck equation and a general
expression for the effective 'temperature' and its average value
the mean-field 'temperature'. We also obtain the generalized Smoluchowski
equation describing the  quasi-equilibrium state. Section 3 is
devoted to  studying the effects of the mean-field 'temperature' on
chemical reaction rates. In section 4 we analyze  the FDT. Finally,
in section 5 we present our main conclusions.

\section{Brownian motion in a field of force}

Let us consider a one dimensional Brownian gas in contact with a 
heat bath at temperature $T_{0}$ \cite{agusti}. Let be $H(\Gamma
)=mu^{2}/2+G(x)$ the Hamiltonian of a particle of mass $m$, where $\Gamma
=(x,u)$ represents a point in the one-particle phase space and $G(x)$ being
the external potential.

The analysis of this Brownian gas is based on thermodynamics through the definition of entropy by means of
  the Gibbs
entropy postulate \cite{groot}%
\begin{equation}
S(t)=-k_{B}\int f\ln \frac{f}{f_{\textnormal{eq.}}}d\Gamma +S_{\textnormal{eq.}},
\label{Gibbs_entropy}
\end{equation}%
where $k_{B}$ is Boltzmann's constant, $f(\Gamma,t)$ is the phase-space distribution 
function,  $S_{\textnormal{eq.}}$ is the  equilibrium
entropy and 
\begin{equation}
f_{\textnormal{eq.}}\sim \exp \left\{ -\frac{H(\Gamma )}{k_{B}T_{0}}\right\} ,
\label{local_maxwelian}
\end{equation}%
is the equilibrium distribution function. Variations in the
probability density $f(\Gamma ,t)$ cause changes in the entropy which can be
obtained from Eq. (\ref{Gibbs_entropy})%
\begin{equation}
\delta S=-\int \left( k_{B}\ln \frac{f}{f_{\textnormal{eq.}}}+\frac{\mu _{\textnormal{eq.}}}{T_{0}}%
\right) \delta fd\Gamma ,  \label{N1}
\end{equation}%
where we have taken into account that $\delta S_{\textnormal{eq.}}=-\int $ $\left( \mu
_{\textnormal{eq.}}/T_{0}\right) \delta fd\Gamma $. Here $\mu
_{\textnormal{eq.}}(x,t)=-G(x)+\mu _{0}$ is the equilibrium chemical potential (mechanochemical potential)
per unit of mass and $\mu _{0}(p,T_{0})$ is the corresponding thermodynamic potential, with $p$ being the pressure. By defining the
nonequilibrium chemical potential 
\begin{equation}
\mu(\Gamma ,t)=k_{B}T_{0}\ln\frac{f}{f_{\textnormal{eq.}}}+\mu_{\textnormal{eq.}},
\label{chemicalpotential}
\end{equation}
the thermodynamic quantity conjugated to the density $f(\Gamma ,t)$,
it is possible to write Eq.(\ref{N1}) in the compact way 
\begin{equation}
T_{0}\delta S=-\int \mu (\Gamma ,t)\delta f(\Gamma ,t)d\Gamma
\label{Gibbs_equation}
\end{equation}%
which constitutes the Gibbs equation in phase space.

A gradient of the chemical potential in phase space (\ref{chemicalpotential}) induces a diffusion process which tends to restore the
equilibrium state. Through this process, the distribution function changes
according to the generalized Liouville equation%
\begin{equation}
\frac{\partial }{\partial t}f(\Gamma ,t)+v(\Gamma ,t)\cdot \nabla _{\Gamma
}f(\Gamma ,t)=-\frac{\partial }{\partial u}J(\Gamma ,t)
\label{generalized Liouville}
\end{equation}%
which defines the diffusion current $J(\Gamma ,t)$ and where $v(\Gamma
,t)=\left( \dot{x},\dot{u}\right) =\left( u,-\nabla G/m \right) $ is the
velocity corresponding to the hamiltonian flow and $\nabla _{\Gamma }=\left(
\nabla ,\partial /\partial u\right) $, with $\nabla =\partial /\partial x$.

From Eq. (\ref{Gibbs_equation}) we can obtain the rate of change of the
nonequilibrium entropy%
\begin{equation}
\frac{dS}{dt}=-\frac{1}{T_{0}}\int \mu (\Gamma ,t)\frac{\partial }{\partial t}f(\Gamma
,t)d\Gamma   \label{entropy_rate}
\end{equation}%
which in combination with Eq. (\ref{generalized Liouville}) and after
partial integration leads to 
\begin{equation}
\frac{dS}{dt}=\frac{1}{T_{0}}\left\langle (-\nabla G)u\right\rangle -\frac{1}{T_{0}}\int J(\Gamma ,t)%
\frac{\partial }{\partial u}\mu (\Gamma ,t)d\Gamma .  \label{Gibbsequation2}
\end{equation}%
Here, the first term on the right hand side of Eq. (\ref%
{Gibbsequation2}) is given by 
\begin{equation}
\left\langle (-\nabla G)u\right\rangle =\int \left[ -\nabla G(x)\right]
uf(\Gamma ,t)d\Gamma   \label{promedio_1}
\end{equation}%
and constitutes the power supplied by the external field which is
dissipated in the system as heat. This contribution to the entropy change comes from the Hamiltonian evolution of the distribution function and therefore cannot be assimilated into the contribution due to diffusion. Thus, Eq. (\ref{promedio_1}) can be
interpreted as the rate of heat exchanged with the surroundings%
\begin{equation}
\frac{dq}{dt}=\left\langle (-\nabla G)u\right\rangle ,  \label{heat_exchange}
\end{equation}%
with $dq$ being the amount of heat released in a time $dt$. In
addition, the second term on the right-hand side of Eq. (\ref%
{Gibbsequation2}) constitutes the entropy production rate due to irreversible
processes $\sigma $ which must be non-negative $(\sigma \geq 0)$ according
to the second law. Therefore, the rate of change of entropy can be written
in the compact way 
\begin{equation}
\frac{dS}{dt}=\frac{1}{T_{0}}\frac{dq}{dt}+\sigma ,  \label{Gibbsequation3}
\end{equation}%
expressing the balance between the exchange of heat with the surroundings
and the entropy generated in the irreversible processes established
in the system. The entropy production contains the current $J(\Gamma ,t)$
and its conjugated thermodynamic force $\left( \partial /\partial u\right)
\mu (\Gamma ,t)$. Following the postulates of nonequilibrium
thermodynamics \cite{groot}, these quantities are related through the
phenomenological law%
\begin{equation}
J(\Gamma ,t)=-\frac{L}{T_{0}}\frac{\partial }{\partial u}\mu (\Gamma ,t),
\label{phenomenological law}
\end{equation}%
where $L$ is the phenomenological coefficient. By using the expression of
the nonequilibrium chemical potential, Eq. (\ref{chemicalpotential}) and
Eq. (\ref{phenomenological law}) one obtains%
\begin{equation}
J(\Gamma ,t)=-\zeta \left(\frac{k_{B}T_{0}}{m} \frac{\partial }{\partial u}+u\right)
f(\Gamma ,t),  \label{phenomenological law2}
\end{equation}%
where we have identified $L/fT_{0}$ as the friction coefficient $\zeta $ of
the Brownian particle ($L/fT_{0}\equiv \zeta/m $). Hence, from the definition
of $\sigma $ given through Eq. (\ref{Gibbsequation2}), along with Eq. (\ref%
{phenomenological law2}) we obtain%
\begin{equation}
\sigma =\frac{m\zeta }{T_{0}}\int \frac{\left[ (k_{B}T_{0}/m)\left( \partial
/\partial u\right) f(\Gamma ,t)+f(\Gamma ,t)u\right] ^{2}}{f(\Gamma ,t)}%
d\Gamma .  \label{entropyproduction}
\end{equation}%
It is worth to emphasize that a stationary nonequilibrium state ($dS/dt=0$)
exist for which 
\begin{equation}
\frac{1}{T_{0}}\frac{d_{st}q}{dt}=-\sigma _{st}.
\label{stationary condition}
\end{equation}%
This corresponds to a stationary state of nonzero entropy production which
differs from the equilibrium state characterized by 
$d_{\textnormal{l.eq}}q/dt=\sigma _{\textnormal{l.eq}}=0$, a condition which is satisfied 
by the local Maxwellian (\ref{local_maxwelian}).

Finally, by substituting Eq. (\ref{phenomenological law2}) into Eq. (\ref%
{generalized Liouville}) we obtain the Fokker-Planck equation
describing the dynamics of the Brownian gas%
\begin{equation}
\frac{\partial }{\partial t}f(\Gamma ,t)=-u\nabla f+\nabla \left[\frac{G(x)}{m}\right]\frac{%
\partial }{\partial u}f+\zeta \left( \frac{k_{B}T_{0}}{m}\frac{%
\partial ^{2}}{\partial u^{2}}f+\frac{\partial }{\partial u}uf\right) .
\label{ten}
\end{equation}

\section{Effective and mean-field 'temperatures'}

At this point, assuming that the velocity $u$ is the fast variable (in
Brownian motion inertia usually constitutes a short lived effect), it is
appropriate to write 
\begin{equation}
f(\Gamma ,t)=\phi _{x}(u,t)\rho (x,t),  \label{eleven}
\end{equation}%
where $\phi _{x}(u,t)$ is the conditional probability density and $\rho
(x,t)=m\int f(\Gamma ,t)du$ is the configurational probability density which
evolves according to%
\begin{equation}
\frac{\partial }{\partial t}\rho (x,t)=-\nabla \left[ m\int uf(\Gamma ,t)du%
\right] .  \label{twelve}
\end{equation}%
Equation (\ref{eleven}) expresses the coupling between the macroscopic 
process triggered by the field $G(x)$ and the microscopic process of the 
momentum relaxation \cite{ross}.
Equation (\ref{twelve}) has been obtained by partial integration of Eq. (\ref%
{ten}) over velocity and thus implicitly defines the diffusion current $%
J(x,t)\equiv m\int uf(\Gamma ,t)du\equiv \rho (x,t)v(x,t)$. This current
satisfies the evolution equation 
\begin{equation}
\frac{\partial }{\partial t}J(x,t)+\zeta J(x,t)=\rho (x,t)\left[ -\nabla
\frac{G(x)}{m}\right] -\nabla \left[ \left( \int u^{2}\phi _{x}(u,t)du\right) \,\rho
(x,t)\right] ,  \label{momentumbalance}
\end{equation}%
obtained from Eq. (\ref{ten}) by multiplying by $u$ an integrating 
by parts with the appropriate boundary conditions. Here, the second moment
of $\phi _{x}(u,t), \int u^{2}\phi _{x}(u,t)du$ is proportional to the thermal energy, \textit{i.e.} the amount of energy of a system necessary for the erratic motion of its Brownian degrees of freedom. This suggests the definition of an effective 'temperature'%
\begin{equation}
\frac{k_{B}T\left( x,t\right) }{m}=\overline{u^{2}}=\int u^{2}\phi
_{x}(u,t)du  \label{N2}
\end{equation}%
in \textit{analogy} with the equipartition theorem \cite{berthier},\cite{granular},\cite{berthier2}. If in
particular, the initial distribution is given by 
\begin{equation}
f(\Gamma ,0)=\rho (x,0)\exp \left\{ -\frac{1}{2}\left( u-v(x,0)\right)
^{2}m/k_{B}T_{0}\right\} ,  \label{initial distribution}
\end{equation}%
the solution of Eq. (\ref{ten}) at later times will have also the same
Gaussian form and is given by (see Ref. \cite{groot} CH. IX, $\S 8$) 
\begin{equation}
f(\Gamma ,t)=\rho (x,t)\exp \left\{ -\frac{1}{2}\left( u-v(x,t)\right)
^{2}m/k_{B}T_{0}\right\} .  \label{N3}
\end{equation}%
In this case the second moment of $\phi _{x}(u,t)$ becomes 
\begin{equation}
\overline{u^{2}}=\left[ \frac{k_{B}T_{0}}{m}+v(x,t)^{2}\right]
\label{secondmoment}
\end{equation}%
which defines the effective 'temperature'%
\begin{equation}
k_{B}T\left( x,t\right) =k_{B}T_{0}+mv(x,t)^{2}.  \label{effectivetemperature}
\end{equation}%
This effective 'temperature' enters the expression of the diffusion current
which, from Eq. (\ref{momentumbalance}) for long times as compared to $\zeta
^{-1}$, reduces to 
\begin{equation}
J(x,t)=-D(x,t)\left[ \nabla \rho (x,t)+\frac{\rho (x,t)}{k_{B}T(x,t)}\nabla
\Phi (x,t)\right] ,  \label{fourteen}
\end{equation}%
where now $\Phi (x,t)=G(x)+k_{B}T(x,t)$ is an effective potential and $%
D(x,t)=(k_{B}T(x,t)/m)\zeta ^{-1}$ is the bare effective diffusion coefficient.
Substitution of Eq. (\ref{fourteen}) into (\ref{twelve}) yields the
generalized Smoluchowski equation 
\begin{equation}
\frac{\partial }{\partial t}\rho (x,t)=\nabla \left\{ D(x,t)\left[ \nabla
\rho (x,t)+\frac{\rho (x,t)}{k_{B}T(x,t)}\nabla \Phi (x,t)\right] \right\} .
\label{nonIsotSmoluch}
\end{equation}

If the conditions are such that the system is in a quasistationary state
characterized by Eq. (\ref{stationary condition}), thus, from Eqs. 
(\ref{heat_exchange}), (\ref{entropyproduction}) and ( \ref{N3}) one obtains 
\begin{equation}
v(x,t)=\zeta ^{-1}\nabla \left[\frac{G(x)}{m}\right],  \label{stationaryvelocity}
\end{equation}%
which implemented in Eq. (\ref{effectivetemperature}) gives%
\begin{equation}
k_{B}T\left( x,t\right) =k_{B}T_{0}+\frac{1}{m}\left[ \zeta ^{-1}\nabla G(x)\right]
^{2}.  \label{stationary temperarture}
\end{equation}

A particular simple case corresponds to the potential $G(x)=-Fx$ for which
Eq. (\ref{stationary temperarture}) reduces to%
\begin{equation}
k_{B}T\left(t\right) =k_{B}T_0+\frac{1}{m}\left( \zeta ^{-1}F\right) ^{2}.
\label{forcetemperature}
\end{equation}

Another interesting particular case corresponds to the presence of thermal
forces. The existence of, for example, an homogeneous temperature gradient
in the bath brings the system to a nonequilibrium state. This
non-homogeneous bath temperature $T_{\textnormal{nh}}(x)$ yields a Fokker-Planck 
equation in which $G(x)=\tilde{\gamma} \ln T_{\textnormal{nh}}$ with 
$\tilde{\gamma}$ a characteristic energy, \cite{mazur} and therefore%
\begin{equation}
k_{B}T\left( x,t\right) =k_{B}T_{\textnormal{nh}}+\frac{1}{m}\left( \gamma \zeta ^{-1}\nabla \ln
T_{\textnormal{nh}}\right) ^{2}  \label{temp_noneq_sys}
\end{equation}

Eqs. (\ref{stationary temperarture})-(\ref{temp_noneq_sys}) show that when
the system is subjected to a large force, the averaged kinetic energy
related to Brownian motion increases, enabling one to define
an effective 'temperature' differing from the corresponding temperature of
the bath. 

In addition, for practical purposes it is convinient to introduce an estimate of the 
'temperature' field $T(x,t)$ which allows to incorporate the effects of the
strong external forces in such a way that the usual techniques of nonquilibrium
statistical physics can be used to calculate the properties of the system. In the 
lowest approximation, this can be done by means of the mean value $\langle T(x,t)\rangle
=\int T(x,t)\rho (x,t)dx$, which we will call the mean-field
'temperature' $T_\textnormal{mf}$:%
\begin{equation}
k_{B}T_\textnormal{mf}(t)=k_{B}\left\langle T_{0}\right\rangle+
\frac{1}{m}\langle \left[\zeta ^{-1} \nabla G(x)\right]^{2}\rangle .  \label{N4}
\end{equation}%
Adopting this mean-field 'temperature', the current (\ref{fourteen}) takes
the form 
\begin{equation}
J(x,t)=-D\left[ \nabla \rho (x,t)+\frac{\rho (x,t)}{k_{B}T_\textnormal{mf}}\nabla G(x)%
\right] ,  \label{sixteen}
\end{equation}%
which, after substituted into Eq. (\ref{twelve}), yields the \emph{%
generalized Smoluchowski }equation 
\begin{equation}
\frac{\partial }{\partial t}\rho (x,t)=D_\textnormal{mf}\nabla ^{2}\rho (x,t)+\zeta
^{-1}\nabla \left[ \rho (x,t)\nabla G(x)\right] ,  \label{seventeen}
\end{equation}%
where $D_\textnormal{mf}=(k_{B}T_\textnormal{mf}/m)\zeta ^{-1}$ is the \textit{mean-field diffusion
coefficient}. At  quasi-equilibrium, the solution of Eq. (\ref%
{seventeen}) 
\begin{equation}
\rho _{\textnormal{qe}}(x)\sim \exp \left[ -\frac{G(x)}{k_{B}T_\textnormal{mf}}\right] ,
\label{eighteen}
\end{equation}%
is characterized by a mean-field thermal energy $k_{B}T_\textnormal{mf}$. In Eq. (\ref%
{eighteen}) we can interpret the mean-field  'temperature' as the
'temperature' for which the configuration probability density $\rho _{\textnormal{qe}}$
of the nonequilibrium system is equal to that given by Boltzmann's
probability distribution formula for a  equilibrium system with
'temperature' $T_\textnormal{mf}$. It is convenient to notice here that the
previous results (\ref{sixteen}), (\ref{seventeen}) and (\ref{eighteen}) are
exact when a constant force is applied on the system, because in this case
Eqs. (\ref{forcetemperature}) and (\ref{N4}) are equivalent.

From the previous analysis, it is plausible to assume that inhomogeneities
caused by the application of forces of different type may introduce similar
effects in the dynamical properties of the system. For example, this will be
the case of diffusion in the presence of a shear flow 
\cite{pre2001},\cite{pre2009},\cite{leporini}.
Finally, it is worth stress that introducing the mean-field 'temperature' $T_{mf}$
is consistent only if $\Delta( m v^2) << k_BT_0$. 
Otherwise, the complete nonhomogeneous nonequilibrium effective 'temperature' $T(x,t)$
has to be considered completely.


\section{Modified chemical reaction rates}


As an application of practical interest of the previous analysis, let us
examine how the application of a large external force on a chemical system
modifies the conditions of chemical equilibrium. In order to perform this
analysis, we will assume that $x$ represents the reaction coordinate and
that the reaction itself can be described as a diffusion process along the
reaction coordinate \cite{agusti2}. For chemical systems $G(x)$ is the free
energy  controlling the reaction.

As in Ref. \cite{agusti2}, the current $J(x,t)$ given through Eq. (\ref%
{sixteen}) can be rewritten in the form 
\begin{equation}
J(x,t)=-D_\textnormal{mf}\,e^{-G(x)/k_{B}T_\textnormal{mf}}\nabla e^{\mu _\textnormal{mf}(x,t)/k_{B}T_\textnormal{mf}},
\label{J(x,t)}
\end{equation}%
where we have introduced the chemical potential 
\begin{equation}
\mu _\textnormal{mf}(x,t)=k_{B}T_\textnormal{mf}\ln \rho +G(x).  \label{mu1}
\end{equation}%
When the height of the energy barrier separating the two minima of the
potential is large compared to thermal energy, a fast relaxation towards the
local minima occurs. Then, the chemical potential becomes a piece-wise continuous function of
the coordinates 
\begin{equation}
\mu _\textnormal{mf}(x,t)=\mu _\textnormal{mf}^{-}(x,t)\Theta (x_{0}-x)+\mu _\textnormal{mf}^{+}(x,t)\Theta
(x-x_{0}),  \label{mu2}
\end{equation}%
with $\mu _\textnormal{mf}^{-}$ and $\mu _\textnormal{mf}^{+}$ referring to the chemical potential
at the left and right wells, respectively. Consequently the probability
density also splits as 
\begin{equation}
\rho (x,t)=\rho _{1}(t)e^{-\left[ G(x)-G(x_{1})\right] /k_{B}T_\textnormal{mf}}\Theta
(x_{0}-x)+\rho _{2}(t)e^{-\left[ G(x)-G(x_{2})\right] /k_{B}T_\textnormal{mf}}\Theta
(x-x_{0}),  \label{rho}
\end{equation}%
here $\rho _{i}(t)=\rho (x_{i},t)$ with $i=1,2$ are the values of the
probability density at the minima, $\Theta (x)$ is the step
function and $x_{j}$ with $j=0,1,2$ are the coordinates of the maximum and
the minima of the potential, respectively.

\begin{figure}[tbp]
{} \mbox{\resizebox*{10.0cm}{!}{\includegraphics{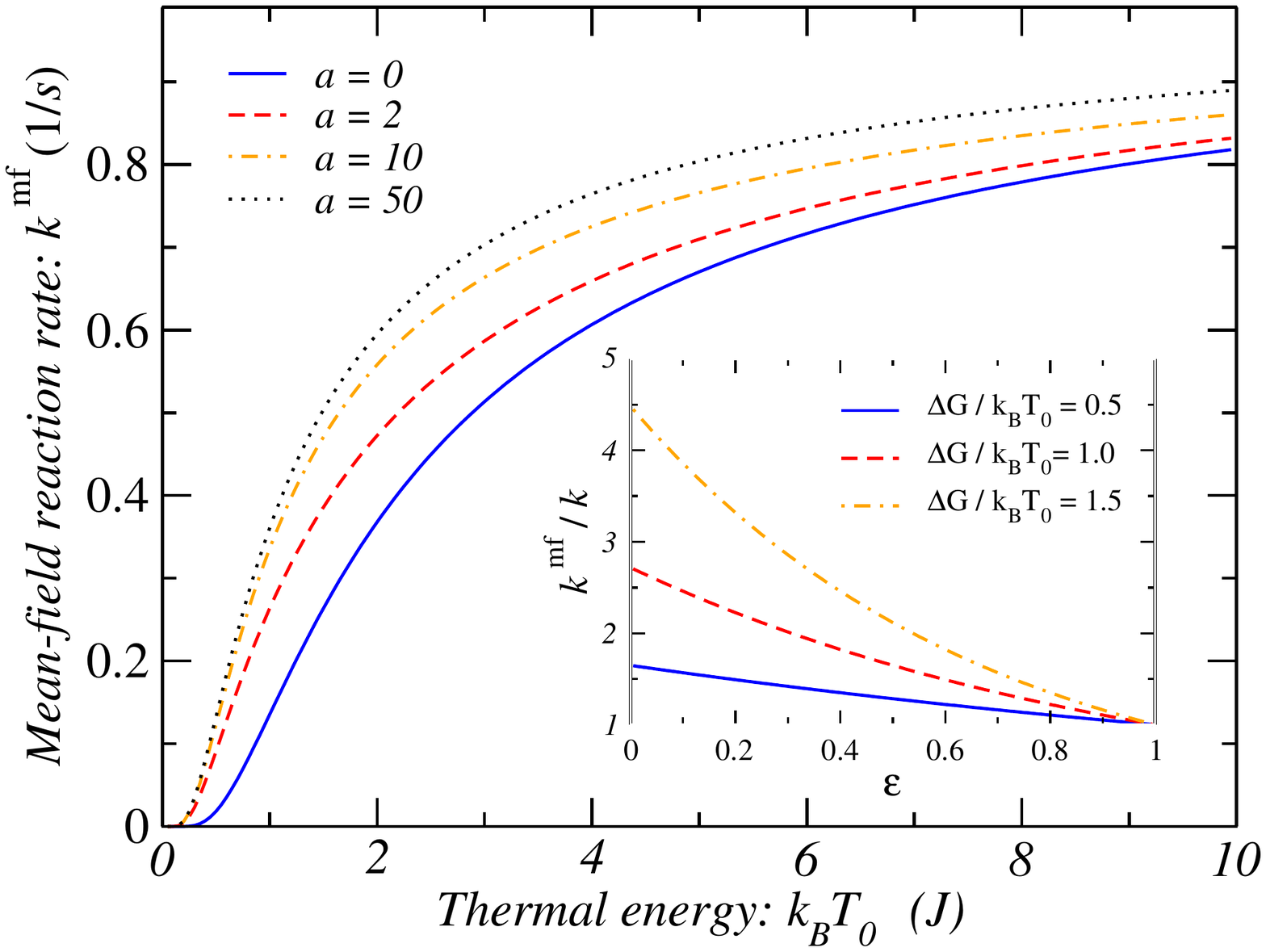}}} {}
\caption{Schematic representation of the mean-field reaction rate $k_\textnormal{mf}$
as a function of the temperature of the bath for different values of the
correction term $a=\frac{1}{m}\langle(\zeta^{-1}\nabla G)^2\rangle$
with $\Delta G=1$ and $D_\textnormal{mf}\protect\sqrt{G^{\prime
\prime}(x_{2,1})|G^{\prime \prime}(x_0)|}/({2 \protect\pi k_B T_\textnormal{mf}})=1$.
The inset shows the ratio $k^\textnormal{mf}/k$ of reaction rates as a function of $%
\protect\epsilon=T_0/T_\textnormal{mf}$ for three different values of 
$\Delta G/k_BT_0 $. When the force associated to the energy $G(x)$ is large the
thermal energy is large enough to increase the mean-field reaction rate. }
\label{f}
\end{figure}

Starting from the mean-field Smoluchowski equation (\ref{seventeen}) and
using relations (\ref{J(x,t)})-(\ref{rho}), it is possible to derive the
following kinetic equation for the concentrations $\rho_{i}$ 
\begin{equation}
\frac{d\rho_{1}}{dt}=-\frac{d\rho_{2}}{dt}=k_{+}^\textnormal{mf}\rho_{2}-k_{-}^\textnormal{mf}\rho_{1},
\label{kinetic1}
\end{equation}%
where the forward and backward reaction rates are given by 
\begin{equation}
k_{+,-}^\textnormal{mf}=D_\textnormal{mf}\frac{\sqrt{G^{\prime \prime }(x_{2,1})|G^{\prime \prime
}(x_{0})|}}{2\pi k_{B}T_\textnormal{mf}}exp\left[ \frac{G(x_{2,1})-G(x_{0})}{k_{B}T_\textnormal{mf}%
}\right] .  \label{rates1}
\end{equation}%
From this equation it follows an interesting result that can be expressed as
the ratio between the mean-field rate $k^\textnormal{mf}$ and the reaction rate when no
external field is applied $k$: 
$k^\textnormal{mf}/k=exp\left( \Delta G/k_{B}T_{0}\right) exp[-\left( \Delta G/k_{B}T_{0}\right) \epsilon ]$
where $\epsilon =T_{0}/T_\textnormal{mf}$ is a measure of the deviation from
thermal equilibrium and $\Delta G=G(x_0)-G(x_{2,1})$ that is a positive quantity
by definition  \cite{jpcm2004},\cite{kegel}. Figure 1 shows $k^\textnormal{mf}/k$
as a function of $\epsilon $. It is clear that the external force shifts
the chemical equilibrium and increases the reaction rates. This 
is particularly interesting in photochemical
reactions because the correction involved in the mean-field
'temperature' is proportional to the energy of the electromagnetic field,
that is to the number of photons associated to the incident light.
Another interesting case is when the chemical reactions take place in a
nonequilibrium medium like alive organisms or chemical reactors in which the
presence of thermal or concentration gradients as well as
hydrodynamic flows modify the chemical equilibrium.


\section{Generalized fluctuation-dissipation theorem}

Once the generalized Smoluchowski equation (\ref{seventeen}) has been
formulated, it is convenient to derive the generalized
fluctuation-dissipation theorem which relates the time derivative of the
correlation function $C_{AF}(t,t^{\prime })$ with the response function $%
R_{AF}(t,t^{\prime })$ of an observable $A(x)$ through the mean-field
'temperature' when at time $t=t_{0}$ the system is perturbed by the external
field $F(x,t)$ \cite{berthier2},\cite{kubo},\cite{ocio}.  As a consequence of this
and in analogy with slow relaxing systems \cite{chetrite}, the
generalized fluctuation-dissipation theorem offers a statistical tool in
order to evaluate the mean-field 'temperature'.

For a sufficiently weak perturbation around the quasi-equilibrium state
characterized by Eq. (\ref{eighteen}) and the mean-field 'temperature' $%
T_\textnormal{mf}$, the response of the system can be expressed in terms of the
deviation of the average value $\langle A(t)\rangle _{F}$ in the presence of
the force field $F$ with respect to the unperturbed case $\langle
A(t)\rangle _{\textnormal{qe}}$: 
\begin{equation}
\langle A(t)\rangle _{F}-\langle A(t)\rangle
_{\textnormal{qe}}=\int_{t_{0}}^{t}R_{AF}(t,t^{\prime })F(t^{\prime })dt^{\prime },
\label{Response}
\end{equation}%
where we have defined 
\begin{equation}
\langle A(t)\rangle _{F}=\int A(x)\rho _{F}(x,t)dx\,\,\,\,\,\,\text{and}%
\,\,\,\,\,\,\langle A(t)\rangle _{\textnormal{qe}}=\int A(x)\rho _{\textnormal{qe}}(x,t)dx.
\label{af}
\end{equation}

To obtain an explicit expression for the left-hand side of Eq. (\ref{af}),
we may use the fact that the perturbed probability density $\rho_{F}$
satisfies the identity 
\begin{equation}
\rho_{F}(x,t)=\int G_{F}(x,t|x^{\prime},t^{\prime})\rho_{F}(x^{\prime
},t^{\prime})dx^{\prime},  \label{CH-K}
\end{equation}
where $G_{F}(x,t|x^{\prime},t^{\prime})$ is the Green function in the
presence of the perturbation. Considering also that the quasi-equilibrium
solution of Eq. (\ref{seventeen}) in the presence of the perturbation $F$
is: $\rho_{\textnormal{qe}}^{F}=\rho_{\textnormal{qe}} Z_{F}(x,t)$, with $Z_{F}(x,t)\equiv
Z[F(x,t)]$ a functional of the perturbation field $F$. Now, after
substituting $\rho_{\textnormal{qe}}^{F}$ in Eq. (\ref{CH-K}) and rearranging terms
one obtains 
\begin{equation}
\rho_{\textnormal{qe}}(x,t)=\int G_{F}(x,t|x^{\prime},t^{\prime})\frac{Z_{F}(x^{\prime
},t^{\prime})}{Z_{F}(x,t)} \rho_{\textnormal{qe}}(x^{\prime},t^{\prime})dx^{\prime}.
\label{CH-K-2}
\end{equation}

Using Eq. (\ref{CH-K-2}) and the fact that $\rho(x,t)=\int G(x,t|x^{\prime
},t^{\prime})\rho(x^{\prime},t^{\prime})dx^{\prime}$ with $G(x,t|x^{\prime
},t^{\prime})$ the unperturbed Green function, we may establish the \textit{%
quasi-equilibrium fluctuation relation} 
\begin{equation}
G_{F}(x,t|x^{\prime},t^{\prime})=\frac{Z_{F}(x,t)}{Z_{F}(x^{\prime},t^{%
\prime })} G(x,t|x^{\prime},t^{\prime}),  \label{FluctuationRelation}
\end{equation}
which expresses the perturbed Green function in terms of the unperturbed
one. Now, using that $\rho_{F}(x,t_{0})=\rho_{\textnormal{qe}}(x,t_{0})$ at $%
t^{\prime}=t_{0}$, then the following relation holds 
\begin{equation}
\rho_{F}(x,t)=\int
G_{F}(x,t|x^{\prime},t_{0})\rho_{\textnormal{qe}}(x^{\prime},t_{0})dx^{\prime}.
\label{rhotilde-rho}
\end{equation}
The substitution of Eq. (\ref{rhotilde-rho}) in Eq. (\ref{af}) leads, after
using the fluctuation relation (\ref{FluctuationRelation}), to the formula 
\begin{equation}
\langle A(t)\rangle_{F}=\int\int A(x)\frac{Z_{F}(x,t)}{Z_{F}(x^{%
\prime},t_{0})} G(x,t|x^{\prime},t_{0})\rho_{\textnormal{qe}}(x^{\prime},t_{0})\,dx
dx^{\prime }.  \label{af-2}
\end{equation}
Here, Eq. (\ref{af-2}) may be written in a form similar to Eq. (\ref{Response}) by
approximating ${Z[F(x,t)]}/{Z[F(x^{\prime},t_{0})]}$ in its series expansion
up to first order in $F$. This operation gives the integral relation 
\begin{equation}
\langle A(t)\rangle_{F} - \langle A(t)\rangle_{\textnormal{qe}} = \int\int A(x)\left[
F(x,t)-F(x^{\prime},t_{0})\right] G(x,t|x^{\prime},t_{0})\rho_{\textnormal{qe}}(x^{%
\prime},t_{0})\,dx dx^{\prime}.  \label{Response-2}
\end{equation}
Recalling now that for the Smoluchowski operator 
\begin{equation}
Z[F(x,t)]=\frac{exp(F_{0}B(x)/k_{B}T_\textnormal{mf})}{\langle
exp(F_{0}B(x)/k_{B}T_\textnormal{mf}) \rangle},  \label{z-smol}
\end{equation}
where we have used $F(x,t)=F_{0}dB(x)/dx$ for convenience and assumed that $%
F_{0}$ is a constant \cite{agusti}. Therefore, using (\ref{z-smol}) in Eq. (%
\ref{Response-2}), for weak perturbations $exp(F_0B/k_BT)%
\sim1-F_0B/k_BT$ and we obtain 
\begin{equation}
\langle A(t)\rangle_{F} - \langle A(t)\rangle_{\textnormal{qe}} = F_0\left[ \frac {1}{%
k_{B}T_\textnormal{mf}(t)}C_{AB}(t,t)-\frac{1}{k_{B}T_\textnormal{mf}(t_{0})}C_{AB}(t,t_{0})\right]
,  \label{Response-3}
\end{equation}
where $C_{AB}(t,t_{0})=\langle A(t)B(t_{0})\rangle-\langle
A(t)\rangle\langle B(t_{0})\rangle$. Assuming now that the quasi-stationary
state of the system varies slowly enough, then we can replace $T_\textnormal{mf}(t)$ by 
$T_\textnormal{mf}(t_{0})$ in (\ref{Response-3}). The resulting expression can then be
written as the integral of the time derivative of the correlation function
which, after being compared with Eq. (\ref{Response}) finally gives 
\begin{equation}
R_{AB}(t,t_{0}) = \frac{1}{k_{B}T_\textnormal{mf}(t_{0})}\frac{\partial}{\partial t_{0}}
C_{AB}(t,t_{0})  \label{Response-4}
\end{equation}
which constitutes the  quasi-equilibrium fluctuation-dissipation 
relation valid for times $t>t_{0}$. This result implies that the
fluctuation-dissipation relation can be generalized to the quasi-equilibrium
state by incorporating the corrections on system's temperature due to the
large potential $V(x)$, \cite{agusti},\cite{jcp2004}. Clearly, this result is fully
compatible with the quasi-equilibrium fluctuation relation (\ref%
{FluctuationRelation}).


\section{Conclusions}

In this paper we have introduced the concepts of nonequilibrium effective
'temperature' $T(x,t)$ and its average value, the mean-field 'temperature' $%
T_\textnormal{mf}$, by using a generalization of the equipartition theorem. This
mean-field 'temperature' parametrizes the  quasi-equilibrium or
metastable states of a nonequilibrium system. This parameter reveals that in the presence of large external forces the thermal energy of the
system increases, therefore promoting the thermal motion of its
corresponding degrees of freedom.

We have shown that at  equilibrium this mean-field 'temperature' reduces
to the bath temperature while at  quasi-equilibrium it contains
quadratic corrections coming from the forces, either Hamiltonian or thermal,
acting on the system. The existence of internal constraints like surface
effects or long-range forces would lead to similar phenomena as the ones
arising from the applied external forces.

Once this mean-field 'temperature' has been introduced, we can describe a
nonequilibrium system as is if it were in equilibrium with a thermal bath at
this $T_\textnormal{mf}$ since it has been proven that
fluctuations in a nonequilibrium system satisfy a generalized
fluctuation-dissipation theorem containing $T_\textnormal{mf}$.

We have illustrated the possible implications in practical situations by
analyzing how external forces modify the chemical equilibrium and the Kramers
reaction rates. We point out that external forces
increase the reaction rates, such as in photochemical reactions or in the presence of nonequilibrium substrates. 

\section*{Acknowledgements}

We acknowledge Prof. J. M. Rubi by let us know the work by J. Ross and P. Mazur, Ref. \cite{ross}. This work was supported by the DGiCYT of Spanish Government under Grant No.
FIS2008-04386 and by UNAM DGAPA Grant No. IN102609. We also thank the
Academic mobility program between the University of Barcelona and the
National Autonomous University of Mexico. This work was finished during a sabbatical stay at the Departament de F\'isica Fonamental of the Universitat de Barcelona (2011).

\newpage

\begin{figure}[tbp]
{} \mbox{\resizebox*{15cm}{!}{\includegraphics{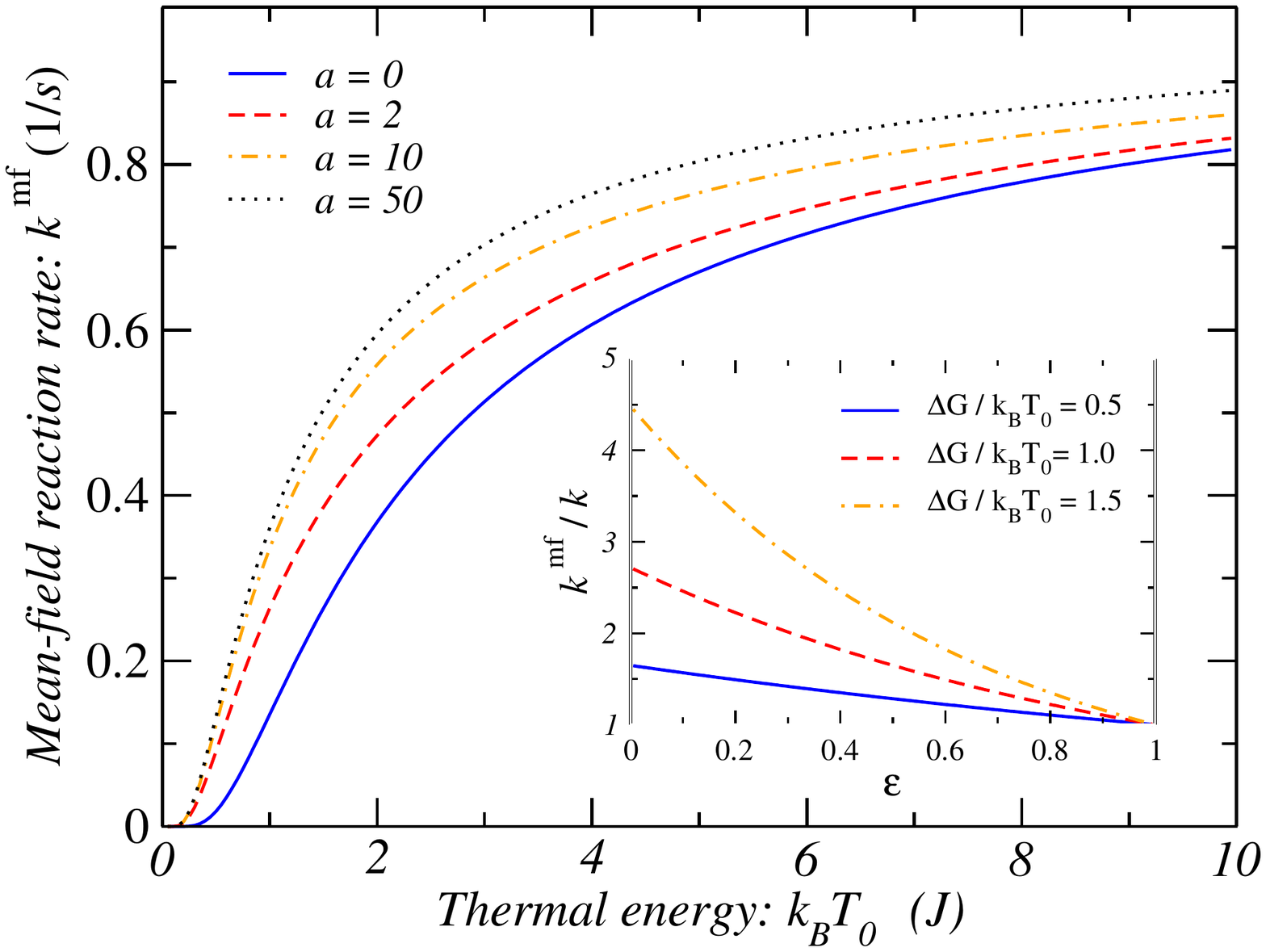}}} {}
\caption{ }
\label{f}
\end{figure}

\end{document}